\documentstyle[12pt]{article}
\newcommand{\be}{\begin{equation}}
\newcommand{\ee}{\end{equation}}
\begin{document}
\vskip20mm
{\large
\begin{center}
{\bf Quantization of the space--time based on a formless finite
fundamental element}
\end{center}
\vskip10mm
\begin {center}
S. B. Afanas'ev

St.-Petersburg, Russia

e-mail:~serg@ptslab.ioffe.rssi.ru
\end {center}
\vskip10mm
\begin{abstract}
The concept of the space (space--time) of the formless finite
fundamental elements (FFFE) is suggested.
This space can be defined as a set
of coverings of the continual space
by non--overlapping simply connected regions of any form and
arbitrary
sizes with some probability measure. The average sizes of each fundamental
element are equal to the fundamental length.
This definition enables to describe correctly
the passage from the space of the
formless finite fundamental elements to the continual space in the limit
of zero value of the fundamental length.
FFFE
space--time functional integral construction is suggested.
A wave function of a separate FFFE and the overall
wave function of a manifold are
introduced. It is shown that many other constructions of the discrete
space--time (the Regge coverings, the lattice space--time etc.) are
the special cases of this space--time.

A vacuum action problem is analyzed. One term of this action is
proportional to the volume of a fundamental element. It is possible to
direct the way for this term to yield the Nambu--Goto action
in consideration the string as one--dimensional excitation
of a number of FFFEs.
Fermionic and bosonic fields
in the space--time of FFFEs are excited states of elements. Space--time
supersymmetry leads to supposition that the
maximal possible number of fermionic
excitations at one FFFE is equal to the number of elements in all space--time.
The compactification in this
space--time means the condition of the neighbour
elements absence in compactificated dimensions.
\end{abstract}
\newpage
\section*{I. Introduction.}

In the classical theory the space--time is a continuum, where the
fundamental elements are points.
In continual geometry all the geometrical objects
are sets of points. Scalar, vector and tensor quantities are the functions
of point coordinates.
Mathematical analysis operations (limits, derivations
etc.) are defined at a point.

In the Dirac quantum mechanics and the
quantum field theory the space--time is also
represented as a continuum. In the classical theory of gravitation
the space--
time is the continuum. But the quantum analysis of the space--time
properties provides some arguments counting in favour of
the
existence of the minimal length, that can be measured by physical methods.
The Heizenberg uncertainty relations application to the process of
small distances measuring yields the inequality~\cite{tred}
\be
(\Delta L)^2 \ge 2\, l_{pl}^2
\ee
where $l_{pl} = (G^{-1} {\hbar}^{-1} c^3)^{1 \over 2} \simeq  10^{-33} sm$.

The analysis of the
space--time properties on the Plank distances leads to the
idea of "space--time foam" [2-3]. The postulate about a fundamental length
($l_f$)
existence is realized in the concept of the quantized (discrete) space--time,
that consists of the fundamental
elements with the finite sizes [4-16].
The lattice space--time with
the fixed lattice is most commonly
 investigated [9-13].

But the space--time with the fixed lattice consideration
leads to several problems. The first and seemingly the
most essential problem is the passage
from the lattice space--time to the continuum in the limit
$l_f \to 0$. The lattice space--time has the power of a countable set.
Any
subdivision of a lattice yields a set with the same power.
Thus in the limit $l_f \to 0$ any lattice space--time
with any subdivision remains a set with the power of a countable
set.
The second, the resulting equations in the lattice space
and other spaces with determined fundamental element form
depend on the form of
a fundamental element.
The third, the
equations in the lattice space are non-invariant under the continual
symmetry operations.

Regarding the quantum ideology all physical quantities don't have
any determined values. Thus if the quantum concept is applied to the
space--time consistently, then it is possible to operate on only
the average sizes
of the elements. In this sense all the determined forms of the
space--time fundamental element
including stereohedra are not consistently quantum
description of the space--time. Consistently quantized space--time can be
formed by the fundamental elements without any determined
forms and sizes only. Nothing but average sizes of
each fundamental element in consistently quantized
space--time can be determined.

Discrete geometry has been developed in the direction of the
formless fundamental
element in the last thirty years.
The Regge calculus~\cite{regge} and the space--time foam idea [2-3]
made the first steps on this way.
In refs.~\cite{yak}
the stereohedra space is investigated. This space fundamental element has
some set of forms. Random lattice field theory is analyzed in~\cite{chiu}.
In ref.~\cite{bal} develops the
topological approach to quantized space--time
calculations.
Quantum configuration
space investigated in~\cite{asht} is the method of quantized
space--time description based on the not--fixed "floating lattice".

In the present work the concept of the space of the formless finite
fundamental elements is suggested. In this concept the problem of
the passage to the continuum
in the limit $l_f \to 0$ and several other problems of the lattice
space--time may be solved.

Notations\\
\noindent $n$-- the dimensionality of a space or a space--time\\
$\eta_{ik}$-- the metric tensor of the plane space or space--time\\
$g_{ik}$--a metric tensor of the Riemannian space or space--time\\
$\{a\}$-- a set of coordinates in the space or the space--time of FFFEs\\
$l_f$-- the fundamental length\\
$l_{pl}$-- the Plank length\\
\newpage
Abbreviations\\
\noindent FFFE -- a formless finite fundamental element\\
FL -- the fundamental length\\
Below $l_f \simeq l_{pl}$ is supposed.

\section*{II. Geometry of the space of the formless finite fundamental
elements.}
Introduce the postulates, that differ the geometry with a formless finite
fundamental
element from the continual geometry.

{\bf Postulate 1. The fundamental length ($l_f$) is a geometrical quantity of
length dimension that means the quantum limit of
measurements accuracy in the space (space--time).

Postulate 2. The space (space--time) consists of the
fundamental elements
that have finite sizes.

Postulate 3.
The formless finite fundamental elements
have average sizes by order fundamental length at
every dimension.
All physical and geometrical quantities are described as fields
defined on a set of FFFEs.

Postulate 4. All physical and geometrical quantities don't depend on
the form of concrete FFFE. They can depend on average geometrical
characteristics of FFFEs only.}

In the text below $n$ is the dimensionality
of the space (or space--time) of FFFEs. This dimensionality is equal to the
dimensionality of the continual space. Introduction of the space of FFFEs
axiomatically allows to construct the mathematical objects
and operations in this space without
consideration of the continual space objects and operations.

On the one hand these postulates are the most probable to be obtained
consistently from analysis of the space--time quantum properties
at the small distances. On the other hand the axiomatical definition of
the quantized
space--time properties could itself leads to the quantization phenomenon
and the quantum field theory as the consequences of the space--time
structure.

The postulates 1 and 2
are identical to the postulates of lattice and
stereohedra
geometries. But the postulates 3 and 4 specifies the geometry with a formless
finite
fundamental element from other geometrical construction of the discrete
space.

The postulate 4 allows to define the space of FFFEs as the set of coverings
of the continual space by any number of non-overlapping simply connected
regions of any form and arbitrary sizes. This set is provided with the
probability measure, i.e. each covering contributes to the space with some
probability. This measure enables the calculations based on this coverings
set (see the sections III, IV). The average values of sizes of FFFEs
are equal to $l_f$, and the average number of FFFEs localized in the
continual space region by the volume $V$ is $N = [V (l_f^n)^{-1}]$.
But the configurations with greatly
different from $l_f$ FFFEs sizes also have the finite probabilities,
for example,
the configurations with one fundamental element that
expands on all the space--time (or investigated manifold),
or the configuration of continual space--time region
itself, i.e. covering this
region by points.
This set of coverings
have the power of continuum.
Therefore limit passage from the space of FFFEs to the
continual space can be
carried out correctly.

In the space of FFFEs the coordinates can be introduced in the region,
consisting of a number of FFFEs. The space coordinates on one
fundamental element don't have a determined meaning in the space of FFFEs and
can be of auxiliary character.
The coordinates introducing on all the set of FFFEs is difficult problem
due to a number of fundamental elements (regions from coverings) on one
manifold is variable. The coordinates can be introduced with sufficient
correctness only on the set of configurations in which all sizes
of all elements are about equal to $l_f$, all $m$--dimensional areas
are about equal to $l_f^m$, and all elements volumes are about equal to
$l_f^n$.

All geometrical operations in the space of FFFEs are determined with
accuracy $O(l_f^k)$. Thus generators of the rotation group in the space of
FFFEs are rotations on a finite angle. Evidently the infinite small
transformation like the ones in the continual space
cannot be the space transformation operations defined
in the space of FFFEs, because the infinite small transformation
doesn't cause any modifications in the set of FFFEs.
The translation group generators are
the translations on a finite distance (by order $l_f$).

This accuracy limit of operations determination
helps to solve the same problems arising in the lattice
space--time consideration.
Thus the lattice Dirac equation is relativistic invariant
with averaging on the continual rotation group only~\cite{darling,shiff}.
In the space--time
of FFFEs this problem is solved at the postulate level, because this
averaging  is the consequence of the postulates 1-4.

The many other constructions of the discrete space--time (i.e. the Regge
coverings, the lattice space--time, the random lattice space--time, the
stereohedra space--time) are the special cases of FFFE space--time with
the special choice of the probability measure. Thus the lattice space--time
is the set of coverings with the probability measure that is equal to zero
for the configurations differ from the coverings by $n$--dimensional cubes
with identical sizes. The random lattice space--time is the set of coverings
with the probability measure that is not equal to zero for the coverings
with the rectangular lattice of the variable step.

The mathematical operations and the physical equations in the space and
the space--time
of FFFEs could be obtained by two methods. The first one is based on
the
known operations and equations of the continual space. The operations and
equations in the FFFE space are the ones for average values,
that are calculated by the method of functional integrals. This method is
considered in the section III.

The second method is the postulative introduction
of operations and equations in the space of FFFEs that requires the
definitions of invariant objects on the set of FFFEs.
These objects depend on the place of
concrete FFFE among the
other elements and average sizes and volume, in the same
time they don't depend on the form and the sizes of concrete FFFE.

Invariant objects defined on each fundamental element must be
the invariants of the
complete space transformation group.
One can note that
the plane space of FFFEs has a specific transformation operation that is
absent in the continual spaces and in the FFFE curved space. This
operation is rearranging of elements. Regarding physics the plane
space--time can be free of particles only,
when with geometrical consideration the properties of the plane space
are identical in all the
space. Therefore any number of elements are able to change their
localization
in any order, and space of FFFE or the manifold of this space is transformed
into itself. In the Riemannian space this operation isn't symmetry
operation due to the coordinate dependence of the connection and different
values of excitations probabilities on different fundamental elements.
Due to this rearranging symmetry the plane space and the Minkowski
space--time of FFFEs are completely
stochastized because any FFFE localization
region isn't exactly determined. Riemannian space and space--time
with particle--like excitations are not stochastized since geometrical and
physical properties is chosen from one element to other.

\section*{III. Functional integral in the space--time of FFFEs.}

In the previous section the calculations method with use of invariant
structures of the FFFE space was discussed.

The other way of obtaining the physical equations and
mathematical operations
in the space and the space--time of FFFEs
is calculations with use of a
continual
(functional) integral.
In agreement with the central idea of
the continual
integral theory the calculation of quantum quantities is the integrating
over all possible configurations of the space (space--time)
of FFFEs (i.e. coverings of the space (space--time)) with into account
the corresponding probability measure taken.

Consider the general construction of a functional integral. In the plane
space it is:
\be
Z = \int {\cal D} V e^{-S(s_i)},
\ee
where
${\cal D} V$ is a measure in the set of coverings,
$S(s_i)$ is the plane space (space--time) vacuum action, $s_i$ is the
set
of element parameters (sizes, areas, volume). Here integrating is over
all coverings of the continual
space (space--time) by non-overlapping simply connected
regions of any
forms and any sizes (see below). Average value of a
function on a separate FFFE
is defined by
\be
\label{func}
<f({\{a\}})> = {{\int {\cal D} V e^{-S(s_i)} f_{\{a\}}(x^i)} \over
{\int {\cal D} V e^{-S(s_i)}}}
\ee
where $f_{\{a\}}(x^i)$
is values of the function $f$ at regions of coverings set
which forms the element $\{a\}$, $f(x^i)$
is a function defined in the continual space (space--time).

In the curved space (space--time) a vacuum functional integral is
\be
Z= \int {\cal D} V {\cal D} g_{ik} e^{-S(s_i,g_{ik})}
\ee
Here $S$ is the curved space (space--time) action.
Full Riemannian space--time action includes particle terms (see in detail
in the section VI). This action is the one of the space--time with excited
states, i.e. vacuum action + action of excitations. Average value of an
operator
in the Riemannian space (space--time) is represented by
\be
<A(\{a\})> = { {\int {\cal D} V {\cal D} g_{ik} {\cal D} \varphi_m
A_{\{a\}} (g_{ik}, \varphi_m, x^i) e^{-S(g_{ik}, \varphi_m, s_i)} } \over
{\int {\cal D} V {\cal D} g_{ik} {\cal D} \varphi_m
e^{-S(g_{ik}, \varphi_m, s_i)} }}
\ee
where $\varphi_m$ is any fields defined in the continual space.
Thus the
functional integrating operation is the one of the averaging over
all configurations that form the space of FFFEs with
the corresponding action. This construction is similar to the functional
integral over surfaces in the Polyakov superstring theory~\cite{pol1}.

Integrating in the functional integral is
over all possible coverings of the continual space
by non-overlapping simply connected regions with
arbitrary sizes and forms.
The application of FFFE functional integrals require the information about
the action $S$. This problem is discussed
in the section IV for the vacuum case.

In the space of FFFEs the wave function of each FFFE could be introduced
(see the remark about the coordinates introducing in FFFE space in the
section II).
This function squared determines the probability of finding the concrete FFFE
in the state with an average localization point $\vec r$, the
volume $V$, total
$m$--dimensional areas $S_m$ and sizes $l_i$. Denote it
$\psi_{\{a\}} ({\vec r}, V, S_m, l_i)$.
Here $\{a\}$ is the set of fundamental element coordinates
in the space of FFFEs.
This wave function squared $|\psi|^2$ is a density of probability in the
set of coverings, i.e. $|\psi|^2 d \sigma$ is the probability of finding the
element with FFFE space coordinates $\{a\}$ in the state with continual
parameters $l_i, S_m, V$. Here $d \sigma$ is a measure in
the set of coverings.
In principle any physical quantities could be found as matrix elements
\be
< A(\{a\}) > = <\psi_{\{a\}} ( {\vec r}, V, S_m, l_i) | \widehat A |
\psi_{\{a\}} ( {\vec r}, V, S_m, l_i) >
\ee
where $A$ is a function in the continual space.
The notation $< A(\{a\}) >$ means
the summing over all regions of coverings set which form the element with
the coordinates set $\{a\}$ in
the space (or space--time) of FFFEs.
Summarized quantities are average values of
$< A >$ on the each configuration element, multiplied on $|\psi|^2$.

The state of the space of FFFEs is a covering of the continual space.
It is also possible to introduce the wave function of a space manifold
or all space wave function (see the discussion of the corrections below):
\be
\label{spw}
\Psi = \prod_{\{a\}} \psi_{\{a\}} ({\vec r}, V, S_m, l_i) +
com(\{\psi_{\{a\}}\})
\ee
This $\Psi$ describes a covering of the continual space by non-overlapping
simply connected regions.
$|\Psi|^2$ is the probability density in the set of coverings. This function
describes the state of the space--time without particles,
i.e. particle--like excited states of FFFE.
However, the vacuum itself has
excited states, where the elements
sizes and localization points differ greatly from the average
values. Functions $\psi_{\{a\}}$ are not independent because they describe
non-overlapping regions. Therefore the expression
(\ref{spw}) contains the second term
that is determined by the commutation relations.

\section*{IV. The Minkowski space--time action.}

The functional integral construction considered in the
previous section requires
the information about the action. In this section the case of
the pure vacuum
is discussed. Regarding physics the case of the vacuum is described
by the plane Minkowski space--time.
Any nonvacuum excitation, including the virtual particle
vacuum polarization and the all space constant fields
leads to arising of the connection
and the curved structure of the space--time.

Suppose that one term of the space--time action is proportional
to the volume of FFFE:
\be
\label{vol}
S_V = \int\limits_{FFFE} \alpha\, \sqrt{-g}\, d V
\ee
Here $d V$ and $\sqrt{-g}$ are continual space values.
Integrating in (\ref{vol}) is over one region from
some covering of the continual space--time. This term is analogous
of "space--time foam" action proportional to the volume~\cite{hawking}.
But
this term
of an action is unsufficient for describing the equilibrium configuration of
the space--time of FFFE. Total actions (\ref{vol}) of all
configurations from the set of coverings are equal.

Consider the problem of the space--time action minimum. On the one hand
as a rule in the method of functional integrating
the minimum of considered action
describes the corresponding classical system (a moving pointlike particle
in the Feynman integral, the
space--time with the classical value of a metric tensor
in the integrating over space--time metrics in quantum gravity etc.).
But in the case of the
space--time the classical system is the continual space--time. It means that
the space--time action must have the minimum in the configuration with infinite
number of fundamental elements which are points.
However, the action minimum in the continual space--time configuration,
and as the consequence the finite value of the probability measure
of this configuration, leads to divergence of some integrals, for example,
the average value of a number of fundamental elements on a continual
manifold. It is seems that the probability measure must be small in the
configurations with small number of fundamental elements.

On the other hand the equilibrium state of the FFFE space--time
is the one with average FFFE sizes at a fundamental length. The
more correct
approach to the equilibrium action problem is to find the fundamental
length, firstly, as the average value of a FFFE size, secondly, as realizing
of some quantized action minimum. It means that the average number of
FFFEs in a manifold is determined by the solution of the action minimum
problem.

The complete expression for the space--time action, meeting this
requirement, must contain other terms besides the volume term. The
possible term is the one proportional to the total $n-1$--~dimensional
area of FFFE. In this supposition the vacuum space--time action is
\be
S = \alpha \sum_i V_i + \beta \sum_i (S_{n-1})_i
\ee
where $\sum_i$ is summarizing over all FFFEs.

The constants in the
action expression can be product of the universal constants only.
Thus
in the four--dimensional space--time
\be
\alpha = A\, \hbar^{-1} G^{-2} c^6
\ee
where $A$ is a numerical factor.

Let us direct the way, on which the Nambu--Goto term of the string action
~\cite{gsw} might be obtained from the space--time action (\ref{vol}).
The expression of the action of a space--time element for this
analysis is required. This action is the average value of an action
$S$ with the functional integrating (\ref{func}) using. This action
is denoted by $S_{FFFE}$:
\be
S_{FFFE} = <S_{\{a\}}>
\ee
or
\be
S_{FFFE} = \int \sqrt{-\eta}\, dx^1 dx^2 dx^3 dx^4
\ee
for the four--dimensional space--time.
This construction also could be introduced axiomatically as the invariant
structure (see the section II).

A string in the space--time of FFFEs can be considered as an excitation
of a number of FFFEs forming one--dimensional space--like curve
(in the meaning of FFFE space--time).
In the own reference frame the action of this
excitation is represented in the form
\be
\label{SFFFE}
S = A\, \hbar^{-1} G^{-2} c^6  \int\limits_{FFFEs}
 \sqrt{-\eta}\, dx^1 dx^2 dl\, d\tau
\ee
where $\tau$ is the own time, $l$ is the own space--like
coordinate of an excitation,
$x^1, x^2$ are the transverse space--like coordinates.
Here integrating is over a set of FFFEs, participated
in the excitation propagation. After integration over
transverse coordinates we might obtain ($\gamma$ is the
two--dimensional metric
tensor determinant):
\be
\label{NG}
S = A\, G^{-1} c^3 \int dl\, d\tau\, \sqrt{-\gamma}
\ee
i.e. the Nambu--Goto action for a string.
In (\ref{NG}) the equality
of the FFFE average size on the one dimension to the fundamental
length is taken into account. This result is not completely
correct due to
the transformation problem
of
the four--dimensional metric tensor determinant
$\eta$ in (\ref{SFFFE}) to the two--dimensional one $\gamma$ in (\ref{NG})
and absence of the correct definition of one--dimensional integrating.
In this concept the $p$--branes are considered as the $p$--dimensional
space--like excitations of FFFEs, and the volume term (\ref{SFFFE})
of FFFE space--time excitation
yields the bosonic term of
$p$--branes action
analogically.

\section*{V. The compactification in the space--time of FFFEs.}

In the concept of the
FFFE space--time the multidimensional space--time with the
motion possible on four dimensions only can be described without
any special compactification procedure. The multidimensional space--time
with average sizes on the higher dimensions by order $l_f$ can be
constructed using the postulates about the absence of nighbour elements for
all ones on all the dimensions without four. These average sizes on the
higher dimensions are $l_f$ despite the configuration with sizes,
which are significantly more $l_f$ on these dimensions
contributes to the space--time structure.

But this compactification description is not satisfying as it requires
the introducing the special postulate. The deeper approach to the
compactification problem is to formulate
the neighbour element absence requirement,
caused space--time action structure analysis or some geometrical
requirement.
Erenfest's investigations about stability
of systems with Coulomb interaction shows that 4--dimensionality of the real
space--time connect with particle action, more precisely, with
the interaction
part of particles action. This way the case $n=4$ of the FFFE space--time
dimensions, on which the motion is possible,
yields the minimum of an action
of $n$--dimensional curved space--time.

\section*{VI. Physical fields in the space--time of FFFEs.}

Fermionic and bosonic fields are excited states of FFFEs. As any quantum
particle, excluding a free particle, has a wave function with different
values
$|\psi|^2$ in different points of the space--time, the space--time with
excitations couldn't be the Minkowski space--time. Different values of
$|\psi|^2$ in different FFFEs violate the Minkowski space--time specific
symmetry under rearranging of any number of FFFEs (see section III).
Therefore two interacting particles in the space--time result in the
curved space--time with changeable curvature. One particle
in all space--time or the uniform vacuum polarization leads to
the particle--like excited
curved space--time with the constant curvature.

Particle--like excitations of the FFFE space--time are finite in each
FFFE of considered manifold. The state
of FFFE with particles excitations in the
approximation of ininteracting particles is
described by a wave function
\be
\psi^{ex}_{\{a\}} = | \{a\}, \psi_i >,
\ee
where $\{a\}$ is a set of coordinates in the FFFE space--time, $\psi_i$
is one-particle wave functions values in this FFFE.

A wave function $|\{a\}, 0 >$
is the sum of all excited vacuum states. In the
classical space--time $|\psi|^2 dV$ is interpreted as a probability of
a particle localization in the volume $dV$. In the FFFE space--time the
interpretation $|\psi_{\{n\}}|^2 dV_{fund}$ is a probability of finding this
FFFE in the excited state with the set of quantum numbers (charges) $\{n\}$.
This probabilities equality means the influence of the
other FFFEs with excited
states on the excited state of this FFFE.  In other words, the space--time
of FFFEs is the self--organizing system.

The states of a manifold of the
space--time of FFFEs is described by an overall
wave function, that could be obtained by summarizing
of each FFFE wave function
with the commutation relations taken into account.
Functional integral construction
in the curved space--time requires to include
particles terms of the space--time action in consideration.

It is to suppose that the number of possible fermionic excitation in one
FFFE is finite. In this case the space--time supersymmetry leads to the
supposition about the equality of a number of possible fermionic excitations
in one FFFE and a number of FFFEs in all the space--time.

\section*{VII. Final remarks.}

Ideologically the suggested concept is consistent geometrical approach
to the physics of fundamental particles and interactions. This concept
may help to solve some problems of lattice space--time geometry due to
more consistent quantum approach to the space--time structure problem.
In this space--time the particles are geometrical objects - excited states
of space--time elements. The superstrings can be constructed as propagating
excitations of space--time elements. At this approach the Nambu--Goto
action term is considered as
a result of this excitation volume space--time action term
analysis.
Consideration of the superstrings as
the excitations of the quantized space--time is the step to the
understanding of the superstrings properties at the Plank distances.
With this superstrings and $p$--branes consideration all these objects
are identical at the Plank distances because the excitation of one element
doesn't have dimension in the sense of FFFE space--time.

But some problems of elementary particle and field theories are not
obvious in this concept, i.e. the appearance and the role of gauge
invariance in the FFFE space--time, appearance of charges of Riemannian
space--time excited states, positing of the cosmological evolution
problems and some others.

In conclusion it is some words about axiomatical introducing of the
quantized space--time (see the section II). Certainly
the author can't be sure that the postulates 1 - 4 are most
correct, complete and minimal system of the quantized space--time
axioms. It is not improbable that the postulate 1 about the fundamental
length existence is the consequence of some other axioms system, and at the
same time the quantized space--time properties are defined by the axioms
introducing the set of FFFEs operationally.
But it is to be noted that on this
way the uncertainty relations, quantum field theory and quantization
phenomenon itself are the consequences of these quantized space--time
axioms system. In particular, it is supposed that the Dirac term of
lagrangian density can come as a consequence of coordinates matrix introducing
~\cite{nc} and
the Weyl structure of
FFFE curved space--time. The Weyl structure of FFFE curved
space--time is connected with the finite accuracy of any
geometrical operations. Thus the vector length at its parallel transport
from one FFFE to other is determined with an accuracy of $l_f$. Therefore
the FFFE curved space--time is the Weyl space--time automatically, and
the Weyl distortion of the Riemannian structure of
this space--time is caused by
quantized structure of the space--time principally.

The author thanks M.S. Orlov,
A.V. Klochkov, E.V. Klochkova, A.V. Sokolov, G.S. Sokolova, D.V. Sokolov,
A.B. Vankov for their
friendly support during the work time, A.A. Amerikantsev,
M.E. Golod for their technical assistance and M.A. Tyntarev for
useful discussions.

\end{document}